# Enhance Ferroelectric Structural Distortion via doping Ca in Epitaxial h-Lu$_{1-x}$Ca$_x$MnO$_3$ Thin Films

**Detian Yang[1,2], Yaohua Liu[3] and Xiaoshan Xu[2,4,*]**

[1] Shanghai Key Laboratory of High Temperature Superconductors, Department of Physics, Shanghai University, Shanghai 200444, China
[2] Department of Physics and Astronomy, University of Nebraska, Lincoln, Nebraska 68588, USA
[3] Second Target Station, Oak Ridge National Laboratory, Oak Ridge, Tennessee 37830, USA
[4] Nebraska Center for Materials and Nanoscience, University of Nebraska, Lincoln, Nebraska 68588, USA

E-mail: xiaoshan.xu@unl.edu

**Abstract**

Unlike proper ferroelectricity, the improper ferroelectricity in multiferroic rare-earth mangnites h-ReMnO$_3$ (Re=La-Lu, Y, Sc) thin films features a unique geometric distortion as the primary order parameter and thus stays immune to the depolarizing field in the ultrathin limit. Here, we have managed to tune and boost ferroelectric geometric distortion of h-LuMnO$_3$ thin films by doping Ca. Compressively strained h-Lu$_{1-x}$Ca$_x$MnO$_3$ (x=0.1,0.2,0.3,0.4,0.5) epitaxial thin films are stabilized over sappire substrates by a h-ScFeO$_3$ buffer layer. When the doping concentration x$\geq$ 0.2, the common substrate-induced interface clamping effect that suppresses improper ferroelectricity of h-ReMnO$_3$ in the subnanometer regime can be eliminated. This work establishes a potential quasi-2D ferroelectric system and suggests a general strain engineering method to enhance improper ferroelectricity of hexagonal mangnites.

## 1. Introduction

Rare-earth manganite ReMnO$_3$ (Re=La-Lu, Y, Sc)[1–5] has constituted one of the most extensively studied oxide platforms in which abnormal orderings and novel properties such as charge ordering, orbital ordering, multiferroicity with strong magnetoelectric coupling, metal-insulator transition and giant magnetoresistance can be achieved and tuned by manipulating the interplay between spin, orbital, charge and lattice degrees of freedom. Therein, hexagonal rare-earth h-ReMnO$_3$ (Re=Y, Ho-Lu, Sc) with smaller rare-earth ions feature colossal flexoelectricity, multiferroicity with magnetoelectric coupling, protected vortex domain structures, topological domain-scaling behaviour and domain walls with tunable conductance and thus promising for room-temperature device ferroelectrics[6–8]. In h-ReMnO$_3$, large oxygen ions arrange into the hexagonal closest-packing ABABAB… structure and the rare-earth ions substitute the oxygen ions in the basal plane every four layers; due to the small size of Mn ion, every two adjacent face-sharing tetragonal sites formed by three continuous oxygen layers along the c axis merge into a bipyramid site and accommodate Mn ion alternatively at A- and B-packing positions. The improper ferroelectricity of multiferroic hexagonal manganites h-ReMnO$_3$ is induced by a structural phase transition of high Curie temperatures[9] and the antiferromagnetic/weak ferromagnetic order emerges independently at low temperatures[10]. As it were, at $T_N$~70-130K, h-ReMnO3 undergoes a magnetic phase transition into the antiferromagnetic state with small c-axis magnetic moments. Therein, the P6_3cm symmetry group adapts a spin-frustrated magnetic order with three adjacent spins mainly forming 120° angles relative to each other in the basal plane as well as canting slightly toward the c-direction due to the Dzyaloshinskii-Moriya interaction[11–13]. At high temperatures, h-ReMnO3 is paraelectric with the centrosymmetric space group $P6_3/mmc$, while as the temperature lowers down below $T_C$~1200-1500K[14], a structural phase transition shifts it into ferroelectric states with a polar space group of P6_3cm through the coupling of strong non-polar K_3 trimerization mode and weak polar $\Gamma_2^-$ mode[15–17]. Such a zone-tripling distortion features the buckling of the Re ion planes and the tilting of the MnO_5 bipyramids, rotating the basal-plane unit cell by 30° and enlarging the basal-plane lattice constant by a factor of $\sqrt{3}$, as shown in Fig.1, where the purple and black frames indicating the unit cell of the paraelectric and ferroelectric phases, respectively. Both the tilting of the MnO_5 and the corrugation of Re-ion layers can be quantified by the order parameter $\vec{Q}$ with two components $Q$ and $\varphi$. On one hand, $Q$ and $\varphi$ correspond to the tilting amplitude and angle of MnO_5, respectively, and on the other $Q$ and $\varphi$ act as the amplitude and phase of the sinusoidal corrugation pattern $Qcos(\vec{k}_{K_3} \cdot \vec{u} - \varphi)$ of Re ions, where $\vec{k}_{K_3}$ is the wavevector of the $K_3$ mode and $\vec{u}$ the position vector of Re ions. The bulking pattern and tilting of MnO_5 bipyramids are indicated by purple curves and white arrows in Fig.1(a) and (b), respectively.

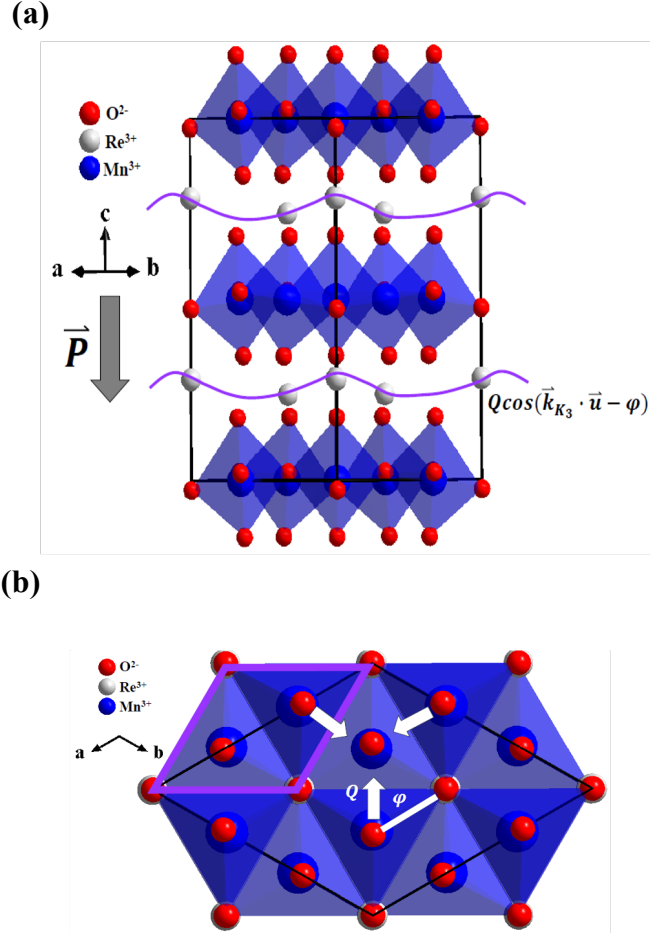

Fig.1. Schematic structure of h-ReMnO3 as T<T_C: (a) along [110] direction; (b) along [001] direction. Both the buckling of Re-planes and tilting of MnO5 bipyramids can be parameterized by two components $Q$ and $\varphi$ of the order parameter $\vec{Q}$ and this ferroelectric distortion results in the rotation of the basal plane unit cell by an angle of 30° and enlarging the basal plane lattice constant with a factor of $\sqrt{3}$, relative to the paraelectric phase with the space group $P6_3/mmc$. The purple and black diamond frame indicate unit cells in the paraelectric and ferroelectric phase, respectively.

The proper ferroelectricity in typical perovskite ferroelectrics such as BiFeO3 [18] and PbZr_{1−x}Ti_xO_3[19], is subject to a so-called critical thickness below which ferroelectricity is substantially quenched if the depolarization field is not fully screened. Evidently, this drawback would hinder the scaling of bulk ferroelectricity down to 2D ferroelectricity of a few atom layers and therefore, deteriorate device miniaturization and energy efficiency. Unlike the proper ferroelectricity where polarization itself is the primary order parameter, the spontaneous polarization in the improper ferroelectrics establishes via the interaction with the primary-order geometric distortion and stays immune to the depolarizing field even in the ultrathin limit. However, in spite of the absence of critical thicknesses as claimed in literature[20,21], a stubborn interface clamping effect suppresses the structural trimerization and makes the improper ferroelectricity unstable in the sub-nanometer ultrathin regime[20].

To date, an effective way to tune and boost the improper ferroelectricity especially in the 2D limit is still unavailbe and highly desirable. Here, we manage to deposit epitaxial h-Lu_{1-x}Ca_xMnO_3 (x=0.1,0.,2,0.3,0.4,0.5) thin films on $\alpha$-Al_2O_3(0001) substrates via predepositing a h-ScFeO_3(001) buffer layer in between. By temperature- and thickness-resolved reflection high energy electron diffraction (RHEED) patterns, we demonstrate that the ferroelectric distortion therein is strengthened with the increasing of the Ca doping concentration x and the absence of the interface clamping effect; as x≥ 0.2, the critical thickness is smaller than 2-3 atomlayers(0.2-0.3unit cells), suggesting a quasi-2D ferroelectric system with a potential out-of-plane polarization.

## 2. Experimental Methods

h-Lu_{1-x}Ca_xMnO_3(001)/h-ScFeO_3(001)/ $\alpha$-Al_2O_3 (0001) epitaxial thin films were grown by pulsed laser deposition

(PLD) in background oxygen pressures 20 mTorr. Both the buffer layer h-ScFeO$_3$ and the h-Lu$_{1-x}$Ca$_x$MnO$_3$ films were grown at 1020 K and annealed at 1100K for 10-30 minutes in the oxygen atmosphere. The KrF excimer laser of the wavelength 248 nm was employed to ablate the targets with a pulse energy of 110 mJ and a repetition rate of 2 Hz. The growth processes were in-situ monitored by a reflection high energy electron diffraction (RHEED) system. The out-of-plane $\theta$-$2\theta$ x-ray diffraction (XRD) and x-ray reflectivity (XRR) were conducted by a Rigaku SmartLab x-ray diffractometer (copper K-α source, X-ray wavelength 1.5406 Å); the film thicknesses were extracted from the XRR data. The in-plane crystal structure and the critical thicknesses for h-Lu$_{1-x}$Ca$_x$MnO$_3$ were studied by analyzing thickness- and temperature-resolved RHEED patterns. For every sample, the thickness-resolved RHEED patterns along the [1 0 0] direction were in-situ recorded at T=1103 K. The temperature-resolved RHEED patterns were recorded with the thickness resolution of about 0.2 unit cells (about 2-3 Å) in the temperature range 570K < T < 1350 K.

## 3. Growth and Preliminary Structural Characterization of h-Lu$_{1-x}$Ca$_x$MnO$_3$ films

(001)-oriented hexagonal manganite thin films can be epitaxially deposited on sapphire (0001)[22] and YSZ (111) substrates [23]. And it has been reported that a buffer layer h-ReMnO$_3$ of higher crystal quality and better flatness can enhance the overall quality of thick h-YMnO$_3$ thin films grown on YSZ substrate[24]. To stabilize Lu$_{1-x}$Ca$_x$MnO$_3$ of the hexagonal structure on rhombohedral sapphire substrates, we first deposit a 0.5-1nm h-ScFeO$_3$ buffer layer. The function of the buffer layer turns out binary: for one thing, it builds up a crystal surface of the favored symmetry and structure compatible with the hexagonal Lu$_{1-x}$Ca$_x$MnO$_3$; for another, it supports compressive strain over the Lu$_{1-x}$Ca$_x$MnO$_3$ film and thus offers an approach to manipulate the thin film's crystal structure, electronic states, and ferroelectricity. Real-time RHEED patterns reveal that impure phases emerge even in the first several atom layers if Ca-doped LuMnO$_3$ is directly deposited on sapphire substrates. Since there is no obvious strain between sapphire substrates and h-ScFeO$_3$ [25] (also see Fig.S1 in the Supplementary Materials) and the in-plane lattice constant of h-ScFeO$_3$[26] is 5% smaller than that of the prototype h-LuMnO$_3$[27], the h-ScFeO$_3$ buffer layer is supposed to exert a compressive strain over Lu$_{1-x}$Ca$_x$MnO$_3$ thin films.

Fig.2 (a) shows typical out-of-plane $\theta-2\theta$ scan of Lu$_{1-x}$Ca$_x$MnO$_3$ thin films, wherein no obvious impurity appears. The epitaxial relation among Lu$_{1-x}$Ca$_x$MnO$_3$ films, the buffer layers and the substrates are summarized in Fig.2(b), derived from RHEED patterns and $\phi$-scan illustrated by the case for h-Lu$_{0.7}$Ca$_{0.3}$MnO$_3$ in Fig.2(c) and Fig.3(a), respectively. Since the buffer layer is of subnanometer, the improper ferroelectric distortion has not established yet[23], so we can assume it is still in the *P6$_3$/mmc* phase. That is why the RHEED rods like (0 1), (0 2), (0 $\bar{1}$) and (0 $\bar{2}$) do not appear. And notice that, both the RHEED patterns of the buffer layers and h-Lu$_{0.7}$Ca$_{0.3}$MnO$_3$ films are labeled based on the reciprocal primitive basis of h-Lu$_{0.7}$Ca$_{0.3}$MnO$_3$ with ferroelectric distortion whose symmetry group is *P6$_3$cm* instead of *P6$_3$/mmc*. As typical examples, in the $\phi$-scans shown in Fig.3(a), (0 3 8) and 5 equivalent peaks of Lu$_{0.7}$Ca$_{0.3}$MnO$_3$ films were measured and demonstrate the six-fold symmetry of Lu$_{0.7}$Ca$_{0.3}$MnO$_3$ films along the out-of-plane direction.

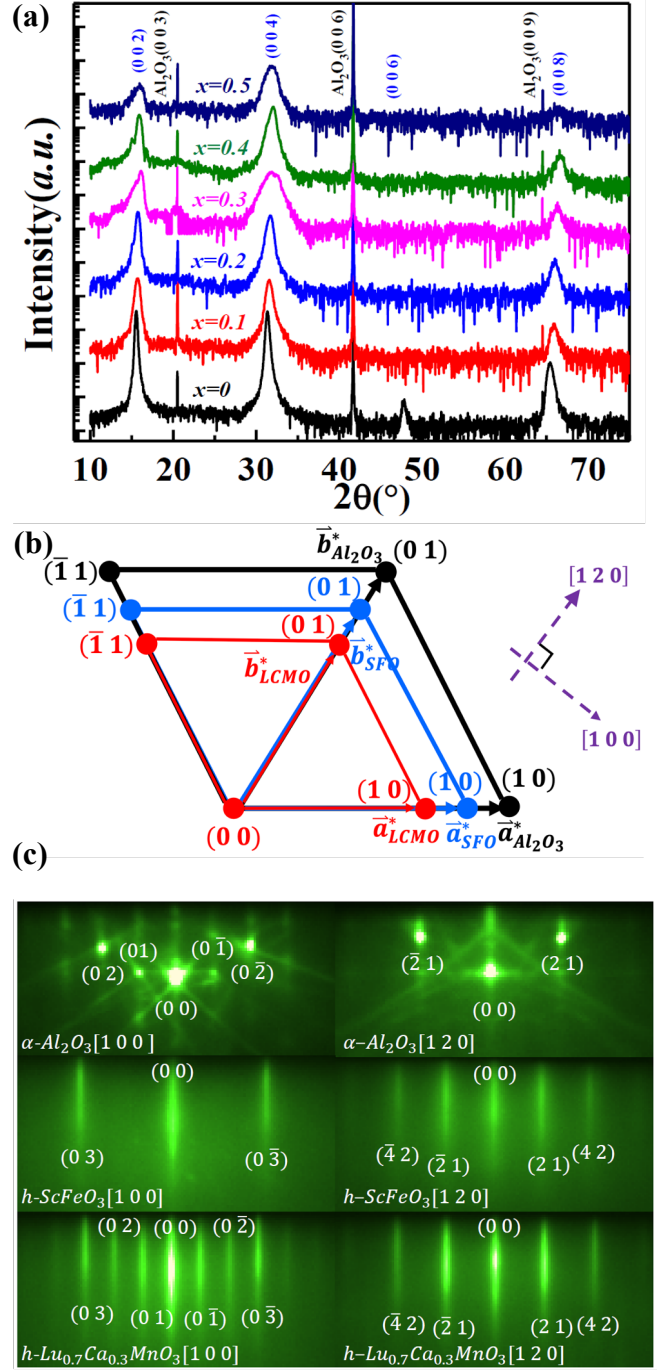

Fig.2. **(a)** $\theta-2\theta$ scans of XRD for h-Lu$_{1-x}$Ca$_x$MnO$_3$ (001)/h-ScFeO$_3$(001)/ α-Al$_2$O$_3$ (0001) samples with x=0 (45 m), 0.1 (21 nm), 0.2 (14 nm), 0.3 (21 nm), 0.4(16 nm), 0.5 (9nm). **(b)** In-plane reciprocal primitive unit cells for the sapphire substrate (black), h-ScFeO$_3$ buffer layer(blue) and h-Lu$_{1-x}$Ca$_x$MnO$_3$ (red); $a_X^*, b_X^*$ denote the in-plane reciprocal lattice constants for material $X$; purple dashed arrows represent two perpendicular e-beam directions. SFO stands for ScFeO$_3$ and LCMO for Lu$_{1-x}$Ca$_x$MnO$_3$. **(c)** Typical RHEED patterns of 30nm Lu$_{0.7}$Ca$_{0.3}$MnO$_3$(bottom panel), 0.5 nm buffer layer(middle panel) and substrates(top panel) along two perpendicular in-plane directions.

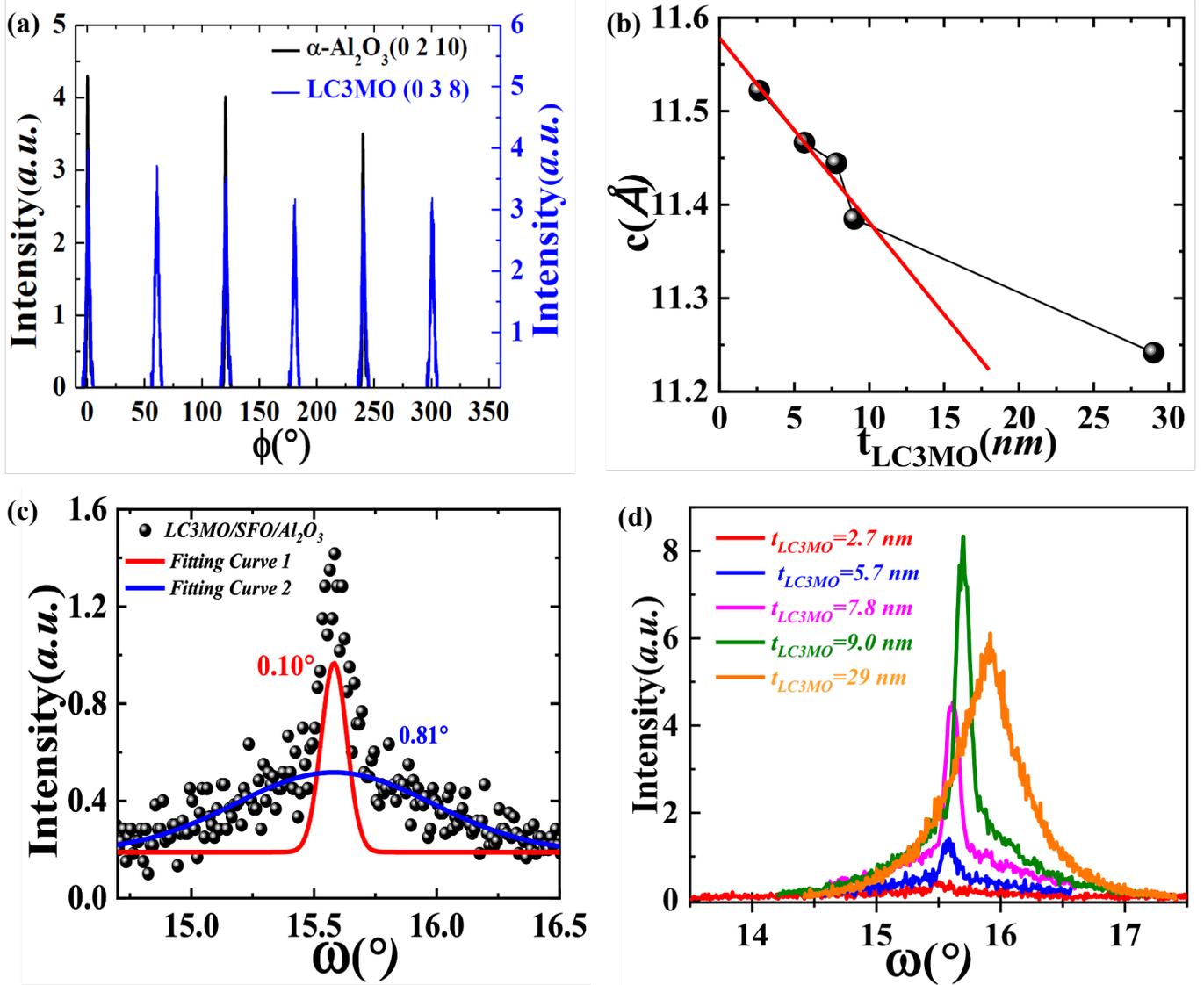

Fig.3. **(a)** $\phi$-scan of h-Lu$_{0.7}$Ca$_{0.3}$MnO$_3$ films and the $\alpha$-Al$_2$O$_3$ substrate for a 29 nm h-Lu$_{0.7}$Ca$_{0.3}$MnO$_3$/h-ScFeO$_3$/$\alpha$-Al$_2$O$_3$ thin film sample; **(b)** Thickness $t_{LC3MO}$ dependent out-of-plane lattice constant c of h-L$_{0.7}$Ca$_{0.3}$MnO$_3$ films grown with the h-ScFeO$_3$ buffer layer; the lattice constant value was derived from (0 0 4) peaks of the samples. **(b)** Thickness $t_{LC3MO}$ dependent out-of-plane lattice constant c of h-L$_{0.7}$Ca$_{0.3}$MnO$_3$ films; the lattice constant values were derived from (0 0 2) peaks of the samples. **(c)** (0 0 4) rocking curve data (dots) and its fittings (lines) of 6-nm Lu$_{0.7}$Ca$_{0.3}$MnO$_3$ films; two sub Gauss curves (red and blue lines) are needed to fit the data well; The numbers indicate the full widths at half maximum of the fitting curves. **(d)** (0 0 4) rocking curves for h-L$_{0.7}$Ca$_{0.3}$MnO$_3$ films of varying thicknesses. Here, LC3MO denotes Lu$_{0.7}$Ca$_{0.3}$MnO$_3$, and SFO for ScFeO$_3$.

## 4. Strain Effects in h-Lu$_{1-x}$Ca$_x$MnO$_3$ films

As discussed in Section 3, our knowledge of pure h-ReFeO$_3$ and h-ReMnO$_3$ thin films grown on sapphire substrates lead to the primary conclusion that Lu$_{1-x}$Ca$_x$MnO$_3$ films would be under compressive strain when the buffer layer is h-ScFeO$_3$. In this section, we directly demonstrate such strain effects by XRD data. In Fig.3(b), the compressive strain in h-Lu$_{1-x}$Ca$_x$MnO$_3$/h-ScFeO$_3$/$\alpha$-Al$_2$O$_3$ is illustrated by the shrinking of the out-of-plane lattice constant c in the case of x=0.3 as the thickness increases. The four data points for thinner samples approximately lie on a red straight line which is the linear fit of them, while the data point at $t_{LC3MO}$=29nm in Fig.3(b) has a lattice constant value of c=11.24 Å even smaller than that of the tensile h-L$_{0.7}$Ca$_{0.3}$MnO$_3$ films as shown in Fig.S2 in the Supplementary Materials. This implies that the strain relaxing feature in Fig.3(b) cannot be fully resorted to the physical strain induced by the buffer layer and the substrate. Fig.3(c) shows the rocking curve of a 6-nm Lu$_{0.7}$Ca$_{0.3}$MnO$_3$ film. At least two Gauss functions with different peak widths are needed to fit the data properly. According to the fitting results, the red subcurve owns a small 0.1° full width at half maximum (FWHM), whereas the blue one has a large FWHM of 0.81°. The peculiar "two-component" feature here can either indicate the emergence of two kinds of crystallites of distinctively different sizes or suggest the misorientation of crystallites by external or internal strains. To examine the potential role of the buffer layer-induced strain effect in such "two-component" feature of the rocking curve, we plot the (0 0 4) rocking curves of multiple-thickness h-Lu$_{0.7}$Ca$_{0.3}$MnO$_3$ films together in Fig.3(d). Clearly, as the thickness $t_{LC3MO}$ is not larger than 9 nm, the sharp component becomes gradually dominant as $t_{LC3MO}$ increases. In view of the expected relief of the buffer layer strain effect with the increasing of the thickness, the "broad component", most likely, is a resultant crystalline misorientation generated the buffer layer strain effect. However, at 29 nm, the "broad component" seems to take control again and the overall rocking curve becomes much broader with an average FWHM of 0.58 °. Although such larger FWHM value might just indicate the deterioration in the crystallinity due to the appearance of more defects, this "anomaly" does echo the abnormally small c value of the same sample as shown in Fig.3(b). In the following section, with RHEED data, we would propose and support that such a deviation at 29 nm reflects the dominant chemical strain induced by the doping effect of Ca as the thickness becomes larger enough.

## 5. Doping-Enhanced Ferroelectric Distortion

As demonstrated in the introduction, the ferroelectric phase transition in h-ReMnO$_3$ would rotate the in-plane primitive unit cell of the paraelectric of *P6$_3$/mmc* phase by 30° and enlarge its in-plane lattice constant by a factor of $\sqrt{3}$. As a result, the in-plane primitive unit cell of the *P6$_3$/mmc* phase denoted by the purple frame in Fig.1(b) reconstructs into a larger primitive unit cell of the *P6$_3$cm* phase highlighted by the black frame. Such a transition induces new RHEED streaks labeled as (0 1), (0 2), (0 $\bar{1}$) and (0 $\bar{2}$) as illustrated in the RHEED patterns of h-Lu$_{0.7}$Ca$_{0.3}$MnO$_3$ along the [1 0 0] direction in Fig.2(c). Thus, by monitoring these streaks, we can investigate the doping effects on the ferroelectric phase transition and the ferroelectricity in h-Lu$_{1-x}$Ca$_x$MnO$_3$ films.

To ascertain the doping effects on the structure and ferroelectricity in h-Lu$_{1-x}$Ca$_x$MnO$_3$ films, for every sample, we in-situ recorded the thickness-resolved RHEED patterns along the [1 0 0] direction at T=1103 K. Fig.4(a)&(b) shows the doping effects of Ca on the relative in-plane lattice constant $\Delta a_{IP}$ of and the relative intensity of (0 1) streak extracted from RHEED patterns for multiple thicknesses. Therein, the data for LuMnO$_3$/YSZ are plotted as a reference. The base line for $\Delta a_{IP}$ is the in-plane lattice constant at t=0; for Lu$_{1-x}$Ca$_x$MnO$_3$ films, it is the in-plane lattice constant of the 0.5 nm h-ScFeO$_3$ buffer layer, but for the LuMnO$_3$ grown on YSZ, it is the in-plane lattice constant of the first several layers of LuMnO$_3$. Clearly, since $\Delta a_{IP}$ for all the Lu$_{1-x}$Ca$_x$MnO$_3$(x=0.1, 0.2, 0.3, 0.4) films increases with the increasing of the thickness as 0<t<15nm, as expected, all the Lu$_{1-x}$Ca$_x$MnO$_3$ films are subject to the compressive strain from the buffer layer h-ScFeO$_3$; while YSZ exerts a tensile strain over LuMnO$_3$ films with an opposite $\Delta a_{IP}$-$t$ feature. In the range 0<t<15nm, all the $\Delta a_{IP}$-$t$ data can be fitted by a general formula $\Delta a_{IP}(t) = y_0 + A_0 e^{-t/t_0}$ with $y_0$, $A_0$ and $t_0$ as the fitting parameters.

To calculate $I_{(0\ 1)}/I_{(0\ 3)}$, we first sum up the intensity along the RHEED streaks of every image to obtain a 1-D intensity data array relative to the pixel numbers along the direction perpendicular to the initial RHEED streaks at every thickness; in doing so, every 2-D RHEED streaks degrades into a peak over the dimension perpendicular to the initial RHEED streaks. Then we locate the peaks corresponding to streaks from streaks (0 1) and (0 3) and extract the maximum value of the peaks as the value of $I_{(0\ 1)}$ or $I_{(0\ 3)}$ to calculate their ratio. By repeating this process for the recorded RHEED images at all thicknesses and combining the results in the end, we achieve a thickness resolved $I_{(0\ 1)}/I_{(0\ 3)}$ data and plot such data for several samples to get Fig.4(b) for the thickness range 0<t<15nm. From Fig.4(b), the relative intensity of $I_{(0\ 1)}$ increases as x increases from x=0.1 to x=0.4. And when x=0.4, with $I_{(0\ 1)}/I_{(0\ 3)}$ >1 for t > 4.5 nm, the (0 1)'s RHEED intensity is even stronger than that of (0 3). Contrary to other cases, the relative intensity $I_{(0\ 1)}/I_{(0\ 3)}$ for x=0.4 keeps increasing as 4.5nm<t<13nm. For a closer look at this, see the [1 0 0]-direction multiple RHEED patterns for Lu$_{0.6}$Ca$_{0.4}$MnO$_3$ films as shown in Fig.4(c). We know the intensity of both (0 1) and (0 3) reflects the crystallinity of the crystal structure of the ferroelectric phase. However, the intensity of streaks induced by the ferroelectric distortion like (1 0) are exclusively in positive correlation with the magnitude of the trimerization of Re ions and tilting of MnO$_5$. So, we assume the ratio $I_{(0\ 1)}/I_{(0\ 3)}$ could reasonably retrieve the strength of the structural distortion of the ferroelectric *P6$_3$cm* phase relative to the paraelectric *P6$_3$/mmc* phase. With such an interpretation, the data of Fig.4(b) demonstrates that the ferroelectric order in h-Lu$_{1-x}$Ca$_x$MnO$_3$ films is enhanced by increasing the doping percent. Because the secondary-order spontaneous polarization develops via the interaction with the first-order geometric distortion and it has been shown that the improper coupling between the polarization and the distortion preserves until the ultra-thin limit[20]. Regardless of the thickness, the polarization is also supposed to be strengthened once we enhance the structural distortion.

*5.1 Strain-Dominant Regime*

Except the boosted ferroelectric distortion, a second intriguing feature in this doping scheme lies in the "rigid" compressive strain from the buffer layer h-ScFeO$_3$ when the doping concentration is not higher than 0.3. As shown in Fig.4(a), as x≤ 0.3, the close $\Delta a_{IP}$-t curves reveal that the strain from the h-ScFeO$_3$ layer releases in a relatively fixed way over the thickness range 0<t<15nm, regardless of the doping level. Whereas, meanwhile, from Fig.4(b), the ferroelectric distortion was gradually reinforced because of the doping of Ca$^{2+}$ ions of a size larger than that of Lu$^{3+}$ ion [28,29]. This is strong evidence that it is the doping effect of Ca that directly strengthens the ferroelectric distortion. However, since we did not observe such enhancement in ferroelectric distortion in tensile-strained h-Lu$_{1-x}$Ca$_x$MnO$_3$ films grown on YSZ, most likely, for x ≤ 0.3 , the enhancement of ferroelectric distortion is a result of the competition between the dominant rigid compressive interfacial strain and the geometric effect of doping(or in other words, the chemical strain due to the doping). Because the compressive strain from ScFeO$_3$ restricts the in-plane lattice constant of Lu$_{0.7}$Ca$_{0.3}$MnO$_3$, the larger Ca ions are restrained in plane to displace more along out-of-plane directions which bolsters the buckling of Re-planes and the tilting of MnO$_5$ with an increase in $Q$. But it seems the enhancement of $Q$ is still limited, because I$_{(0\ 1)}$/I$_{(0\ 3)}$ <1 as x≤ 0.3 as shown in Fig.4(b).

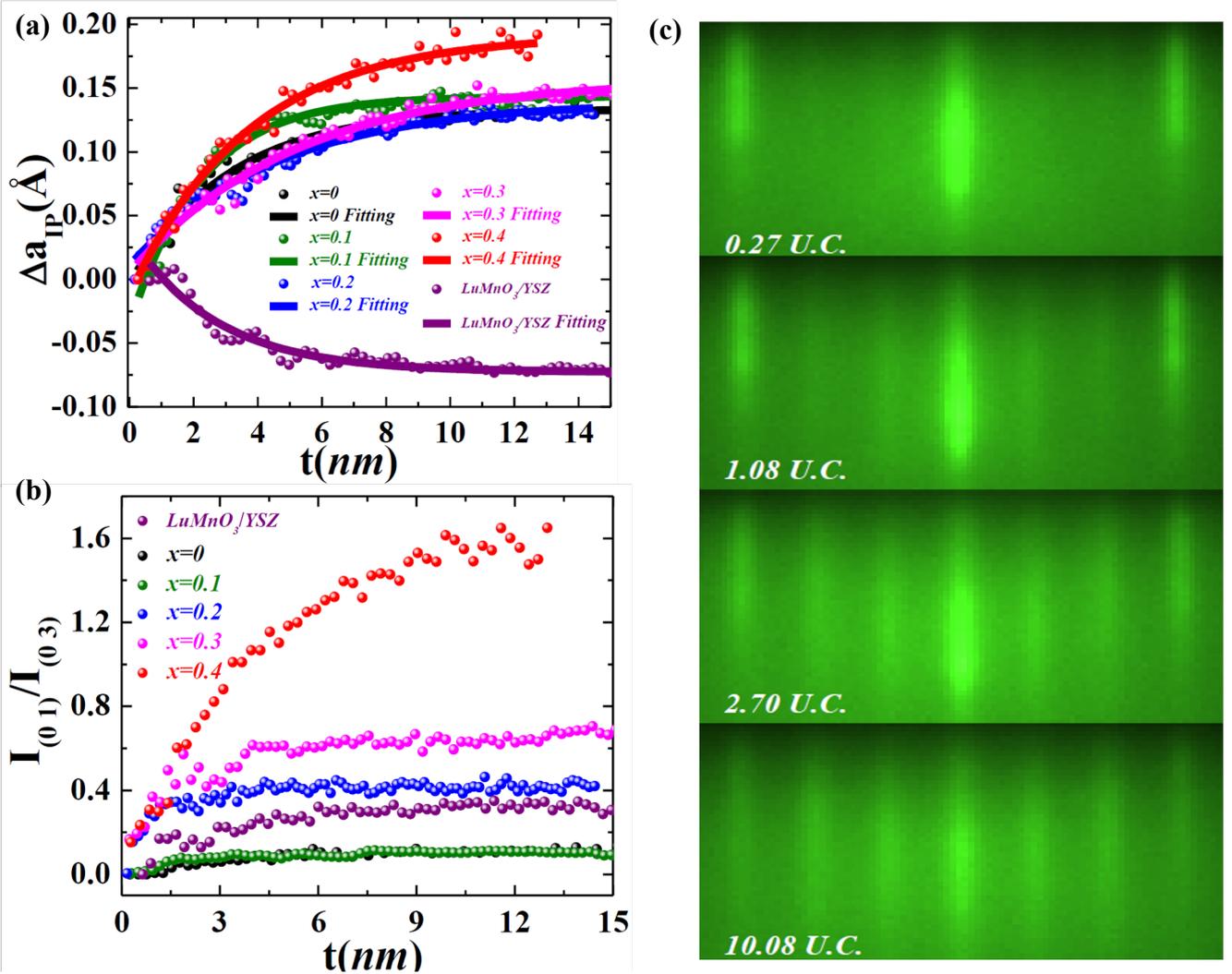

Fig.4. Doping effects on the strain effects and the ferroelectric distortion. **(a)** Thickness t dependent relative in-plane lattice constants $\Delta a_{IP}$ (balls)and their fittings (lines) of h-Lu$_{1-x}$Ca$_x$MnO$_3$ /h-ScFeO$_3$/ α-Al$_2$O$_3$ samples with x=0 (black), 0.1 (olive), 0.2 (blue), 0.3 (magenta), 0.4(red); the base line for $\Delta a_{IP}$ is the in-plane lattice constant at t=0; as 0<t <15nm, all groups of data here can be fitted well by the function $\Delta a_{IP}(t) = y_0 + A_0 e^{-t/t_0}$ with $y_0$, $A_0$ and $t_0$ as the fitting parameters. **(b)** Thickness t dependent relative RHEED intensity of the RHEED streak (0 1) of h-Lu$_{1-x}$Ca$_x$MnO$_3$ /h-ScFeO$_3$/ α-Al$_2$O$_3$ samples with x=0 (black), 0.1 (olive), 0.2 (blue), 0.3 (magenta), 0.4(red); I$_{(0\,1)}$ and I$_{(0\,3)}$ denote the maximum RHEED intensity of (0 1) and (0 3) RHEED streaks; see the text for the details. The purple balls and line here denote the data and its fitting of LuMnO$_3$/YSZ . **(c)** RHEED patterns of h-Lu$_{0.6}$Ca$_{0.4}$MnO$_3$ films with multiple thicknesses along the [1 0 0] direction at T=1103 K.; the corresponding thickness for each image is labeled on the bottom left.; U.C. stands form unit cell.

*5.2 Doping-Dominant Regime*

Fig.4(a) shows that the strain relaxes faster in the $\Delta a_{IP}$-t curve for the $x = 0.4$ case than others. This suggests that, as more larger Ca$^{2+}$ ions replace smaller Lu$^{3+}$ ions the doping effect surpasses the strength of the interfacial strain. Based on the analysis in the strain-dominant regime, in the h-Lu$_{0.6}$Ca$_{0.4}$MnO$_3$ film, both strain relaxing and the Ca doping contribute to the growing of the in-plane lattice constant. The strong geometric doping effects for the x=0.4 case is revealed by the distinguishingly strong relative intensity of the RHEED streak (0 1) with I$_{(0\ 1)}$/I$_{(0\ 3)}$ >1 once t>3.5nm. This means, a larger Q, or the strong tilting of MnO$_5$ and buckling of Lu/Ca-planes achieved by doping Ca, increases the in-plane lattice constant while presumably decreasing the out-of-plane lattice constant.

As proved by the $\Delta a_{IP}$-t curves for x≤ 0.3 cases in Fig.4(a), the physical strain effect from the buffer layer is effectively restrained to a thickness range of 0-13nm. Except the high doping level case of x=0.4, the doping effects are supposed to dominate in the thicker regime even when x≤ 0.3 once the physical strain has been released. To verify this explicitly, we plot $\Delta a_{IP}$-t and I$_{(0\ 1)}$/I$_{(0\ 3)}$-t of two cases x=0.1 and x=0.3 with larger thickness range in Fig.6. Apparently, for t>13 nm, the in-plane lattice constants of both samples are still increasing with the thickness, despite that the slope is much smaller than those in the range t<13 nm. We believe in this larger thickness range, it is mainly the doping, or in other words, the chemical strain from Ca$^{2+}$ ions, that extend in-plane lattice constants much slowly but continuously with the thicknesses. This is supported by the distinctive increasing behaviors of I$_{(0\ 1)}$/I$_{(0\ 3)}$-t for both samples once t >13 nm. In the thicker range (t >13 nm), as the x=0.4 case above, the chemical strain of doped Ca$^{2+}$ boosts the ferroelectric distortion Q which enlarge the in-plane lattice constant and shrink the out-of-plane one. This point explains away the seemingly abnormal data a $t_{LC3MO}$=29nm in Fig.3(b)&(d).

## 6. Absence of Interface Clamping Effect and Quasi-2D Ferroelectricity

In last section, with thickness-resolved RHEED and XRD data, we demonstrate that the doping of Ca could effectively enhance the geometric distortion that leads to improper ferroelectricity in hexagonal manganites. Here, by temperature- and thickness-resolved RHEED patterns, we further exhibit the absence of the interface clamping effect in h-Lu$_{1-x}$Ca$_x$MnO$_3$/h-ScFeO$_3$/$\alpha$-Al$_2$O$_3$ as $x \geq 0.2$. That is, the critical thickness of the ferroelectric distortion is smaller than 2-3 atom layers (0.2-0.3 unit cells), which suggests a potential quasi-2D mangnites-based ferroelectric system with an out-of-plane polarization.

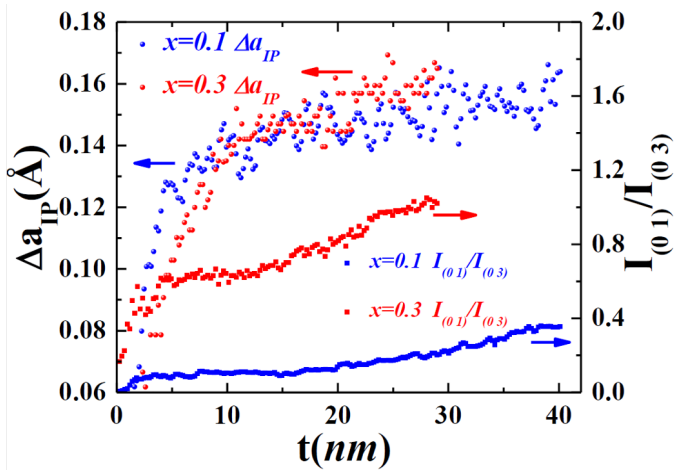

Fig.6. Doping effects in h-Lu$_{0.9}$Ca$_{0.1}$MnO$_3$ and h-Lu$_{0.7}$Ca$_{0.3}$MnO$_3$ films with the larger thickness range: Thickness *t* dependent relative in-plane lattice constants $\Delta a_{IP}$ and relative intensity of the RHEED streak (0 1) of h-Lu$_{0.9}$Ca$_{0.1}$MnO$_3$ and h-Lu$_{0.7}$Ca$_{0.3}$MnO$_3$ films.

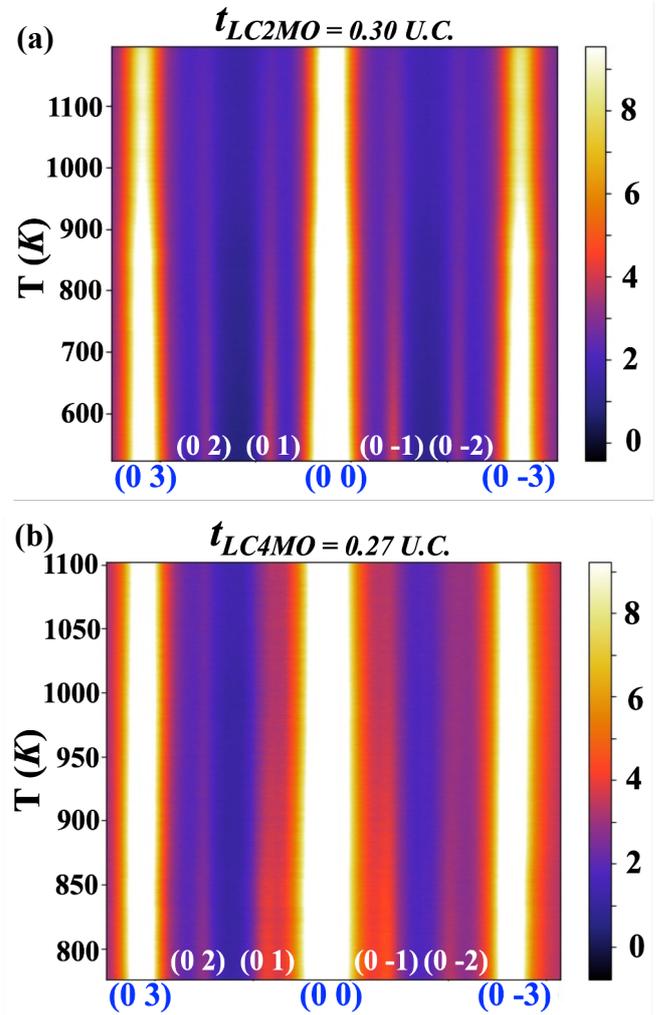

Fig.7. Temperature-resolved RHEED patterns along the [1 0 0] direction at critical thicknesses for **(a)** h-Lu$_{0.8}$Ca$_{0.2}$MnO$_3$ /h-ScFeO$_3$/ $\alpha$-Al$_2$O$_3$; **(b)** h-Lu$_{0.6}$Ca$_{0.4}$MnO$_3$/h-ScFeO$_3$/ $\alpha$-Al$_2$O$_3$. Here, LC2MO and LC4MO stand for h-Lu$_{0.8}$Ca$_{0.2}$MnO$_3$ and h-Lu$_{0.6}$Ca$_{0.4}$MnO$_3$, respectively. And $t_{LC2MO}$ and $t_{LC4MO}$ labeled on the top of the diagrams denote the samples' thicknesses. U.C. denotes unit cell.

PLD is well-known for its capability of the layer-by-layer growth with the atomic-level control[30,31]. With such a capability, we can monitor the thickness dependence of critical thickness and Curie temperature with a resolution of down to one atomic layer. To accurately determine the critical thickness of h-Lu$_{1-x}$Ca$_x$MnO$_3$/h-ScFeO$_3$/$\alpha$-Al$_2$O$_3$ (x=0, 0.1,0.2,0.3,0.4) and h-LuMnO3/YSZ, we recorded the temperature-resolved RHEED patterns along the [1 0 0] direction after the deposition of every layer. The resolution of the thickness, or the thickness of every layer can be controlled by adjusting the corresponding laser shot number. Specifically, we recorded the temperature-resolved RHEED patterns of h-Lu$_{1-x}$Ca$_x$MnO$_3$/ h-ScFeO$_3$/$\alpha$-Al$_2$O$_3$ (x=0, 0.1,0.2,0.3, 0.4) and h-LuMnO3/YSZ with the thickness resolution of about 0.2 unit cells (about 2-3 Å) in the temperature range 570K < T< 1350 K. The critical thickness here is defined as the smallest thickness at which the RHEED streaks (0 1), (02), (0, $\bar{1}$), (0, $\bar{2}$) appear in the temperature range (500-570K)< T< 1350 K. The temperature-resolved RHEED patterns for h-Lu$_{1-x}$Ca$_x$MnO$_3$/ h-ScFeO$_3$/$\alpha$-Al$_2$O$_3$ (x=0, 0.1,0.2,0.3,0.4) and h-LuMnO3/YSZ at the critical thickness were illustrated by the cases x=0.2 and x=0.4 as plotted in Fig.7(For more data, see Fig.S3 in Supplementary Materials). Clearly, for every case, we can unequivocally see the (0 1) streak in the chosen temperature range. At the top of each diagram, the critical thickness is labeled. Around the critical thicknesses, the temperature dependence of $I_{(0\;1)}/I_{(0\;3)}$ for all the samples measured were extracted from the temperature-resolved RHEED patterns and plotted in Fig.8(a)-(c). In the two types of h-LuMnO$_3$ films, the gradual emergence of the ferroelectric distortion with the layer-by-layer growth (here each layer is as thick as about 0.20 U.C.) can be clearly identified by comparing the $I_{(0\;1)}/I_{(0\;3)}$ magnitudes of different thicknesses in Fig.8 (a) and (b). And Fig.8(c) shows, up to x=0.1, the interface clamping effect is probably still working over a range of 0.44 unit cells, but as x ≥ 0.2, the ferroelectric distortion already sets up even in the 2-3 atom layers. Besides, as x increases from 0.2 to 0.4, the $I_{(0\;1)}/I_{(0\;3)}$ become larger with similar thicknesses. This again verifies the doping-enhancement of ferroelectric distortion even in the quasi-2D case. The doping concentration dependent critical thickness is shown in Fig.8(a). The critical thickness for h-LuMnO$_3$/YSZ thin films with a tensile strain reads as about 0.79 unit cells (about 1 nm) close to the value reported earlier[20], while the growth of the same h-LuMnO$_3$ film on the h-ScFeO$_3$ buffer layer with a compressive strain tends to reduce the critical thickness to 0.59 unit cells. The h-ScFeO$_3$ buffer layer seems to be able to weaken the interface clamping effect even without doping. Since 0.2-0.3 unit cells of h-ReMnO$_3$ include about 2-3 atom layers. The improper ferroelectric distortion of h-ReMnO$_3$ cannot establish below such a size considering the feature of $K_3$ mode . This means, once x ≥ 0.2, we achieve the ferroelectric distortion of 2-3 atom layers. The lattice constants employed to estimate the size of the samples' unit cells were derived from the $\theta$-$2\theta$ XRD scans. Hence, by utilizing atomic-level growth control and the highly surface-sensitive RHEED technique, we proves the ferroelectric distortion can be greatly enhanced by doping Ca in h-LuMnO$_3$. And with the increasing of the doping level, the interfacial clamping effects can be weakened and removed completely once the doping concentration reaches x=0.2.

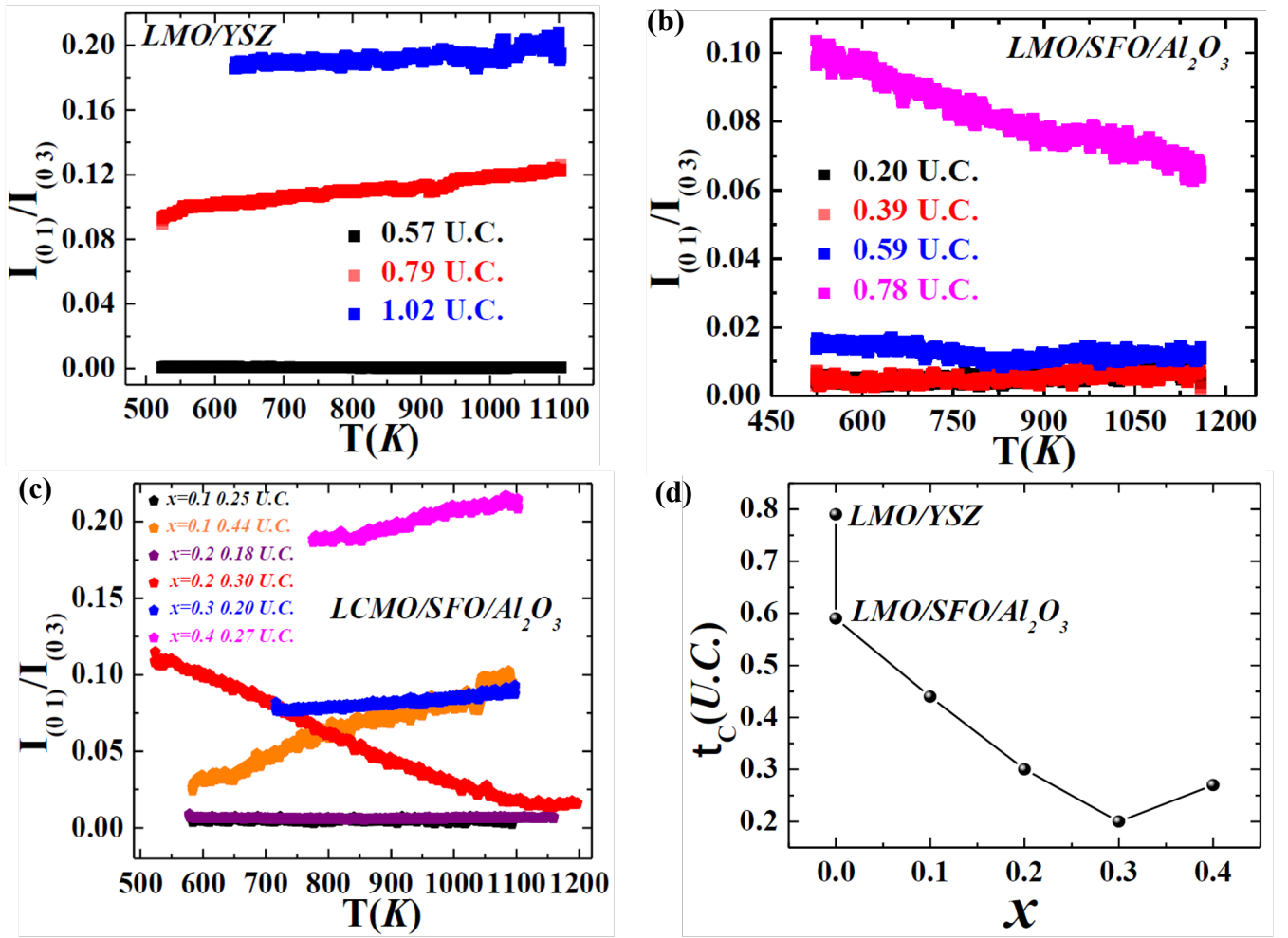

Fig.8. **(a)** Temperature dependent I$_{(0\ 1)}$/I$_{(0\ 3)}$ for h-LuMnO$_3$/YSZ at the thicknesses around the critical thickness $t_C$=0.79 U.C..**(b)** Temperature dependent I$_{(0\ 1)}$/I$_{(0\ 3)}$ for h-LuMnO$_3$/ h-ScFeO$_3$/$\alpha$-Al$_2$O$_3$ at the thicknesses around the critical thickens $t_C$=0.59 U.C.. **(c)** Temperature dependent I$_{(0\ 1)}$/I$_{(0\ 3)}$ for h-Lu$_{1-x}$Ca$_x$MnO$_3$/ h-ScFeO$_3$/$\alpha$-Al$_2$O$_3$ at the thicknesses around the critical thicknesses $t_C$=0.44 U.C., 0.30 U.C., 0.20 U.C. and 0.27 U.C. fore x=0.1. 0.2. 0.3. and 0.4, respectively. **(d)** Doping concentration *x* dependent  critical thickness $t_C$; there in, the two points at x=0 are h-LuMnO$_3$ films grown on the YSZ substate (upper one) and on the sapphire substrate (lower one)with the h-ScFeO$_3$ buffer layer. In this diagram, U.C. stands for unit cell; LMO, SFO, and LCMO stand for h-LuMnO$_3$, h-ScFeO$_3$ and h-Lu$_{1-x}$Ca$_x$MnO$_3$, respectively.

## 7. Conclusions

In conclusion, we have deposited epitaxial h-Lu$_{1-x}$Ca$_x$MnO$_3$ (x=0.1,0.,2,0.3,0.4,0.5) thin films on $\alpha$-Al$_2$O$_3$(0001) substrates by intercalating a h-ScFeO$_3$(001) buffer layer in between. With XRD and RHEED data, we systematically studied the physical strain effects from the buffer layer and the chemical strain effects from the doping of Ca. The Ca dopants was proved to effectively enhance geometric distortions which leads to the improper ferroelectricity. And once the doping ratio x ≥ 0.2, the ubiquitous interface clamping effect that destabilizes the ferroelectric distortions in hexagonal mangnites can be eliminated, which suggests a 2D ferroelectric system with an out-of-plane polarization. This work establishes a general strain engineering method to enhance improper ferroelectricity and stabilize the 2D ferroelectricity in hexagonal mangnites.

## Acknowledgements

The authors acknowledge the primary support from the National Science Foundation (NSF) through EPSCoR RII Track-1: Emergent Quantum Materials and Technologies (EQUATE), Award No. OIA-2044049. The research was performed in part in the Nebraska Nanoscale Facility: National Nanotechnology Coordinated Infrastructure and the Nebraska Center for Materials and Nanoscience, which are supported by the NSF under Grant No. ECCS-2025298, and the Nebraska Research Initiative.

# Supplemental Materials for "Enhance Ferroelectric Structural Distortion by doping Ca in Epitaxial h-Lu$_{1-x}$Ca$_x$MnO$_3$ Thin Films"

Detian Yang[1,2], Yaohua Liu[3] and Xiaoshan Xu[2,4,*]

[1] Shanghai Key Laboratory of High Temperature Superconductors, Department of Physics, Shanghai University, Shanghai 200444, China
[2] Department of Physics and Astronomy, University of Nebraska, Lincoln, Nebraska 68588, USA
[3] Second Target Station, Oak Ridge National Laboratory, Oak Ridge, Tennessee 37830, USA
[4] Nebraska Center for Materials and Nanoscience, University of Nebraska, Lincoln, Nebraska 68588, USA

# 1. The interfacial clamping effect of h-ScFeO$_3$ free from the thickness effect

As already discussed in Section 3 in the main text, the h-ScFeO$_3$ buffer layer plays a key role in the smooth growth of compressively strained h-Lu$_{1-x}$Ca$_x$MnO$_3$ films. To verify that the "rigid" compressive strain exclusively originates from the interfacial clamping exerted by the h-ScFeO$_3$ surface, we grew a h-Lu$_{0.7}$Ca$_{0.3}$MnO$_3$ /h-ScFeO$_3$/ $\alpha$-Al$_2$O$_3$ sample with a 10 nm h-ScFeO$_3$ buffer layer and compare it with the sample with the buffer layer h-ScFeO$_3$ of a thickness of 0.5nm. The results were arranged in Fig.S1. From Fig.S1(a), evidently, the thickness behaviors of $\Delta a_{IP}$ in the two cases match each other quite well, indicating that the thickness of h-ScFeO$_3$ $t_{SFO}$ has no influence on the compressive strain on h-Lu$_{0.7}$Ca$_{0.3}$MnO$_3$ and suggest that such strain effects are purely built up by the interfacial clamping effect between interfacial ions of h-ScFeO$_3$ and h-Lu$_{1-x}$Ca$_x$MnO$_3$. Fig.S1(b) shows that the relative in-plane lattice constant of h-ScFeO$_3$ stays constant as it becomes thicker. This confirms that the interface between h-ReFeO$_3$ and sapphire substrates accommodates no strain[1].

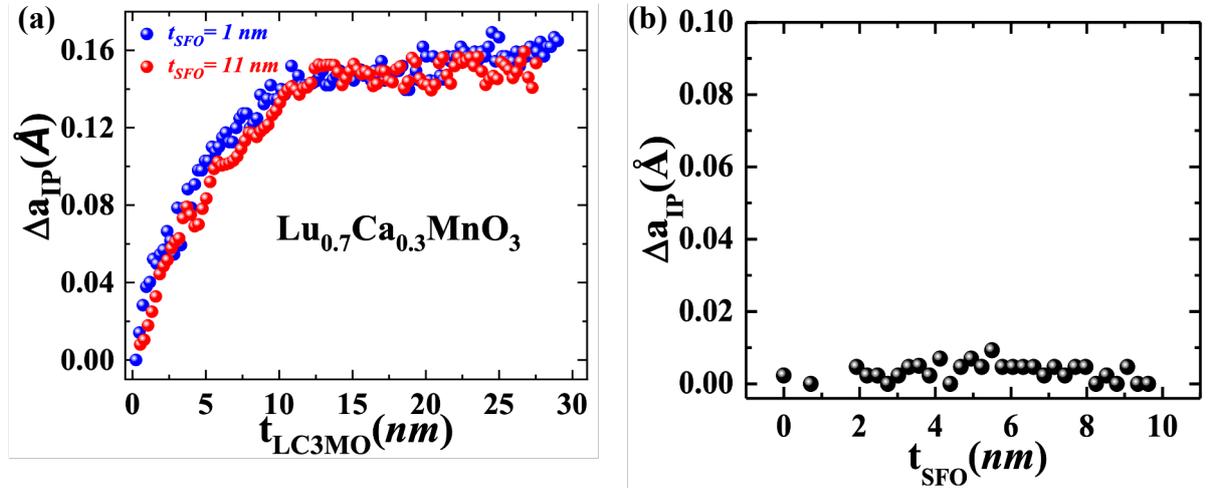

Fig.S1. **(a)** Thickness $t_{LC3MO}$ dependent relative in-plane lattice constants $\Delta a_{IP}$ of h-Lu$_{0.7}$Ca$_{0.3}$MnO$_3$ films with a 0.5 nm (blue balls) and 10 nm (red balls) h-ScFeO$_3$ buffer layers; the base line for $\Delta a_{IP}$ is the in-plane lattice constant at $t_{LC3MO}$=0. **(b)** Thickness $t_{SFO}$ dependent relative in-plane lattice constants $\Delta a_{IP}$ of h-ScFeO$_3$; the base line for $\Delta a_{IP}$ is the in-plane lattice constant at $t_{SFO}$ =0. Here, LC3MO denotes Lu0.7Ca0.3MnO3, and SFO for ScFeO3.

## 2. On the growth of h-Lu$_{1-x}$Ca$_x$MnO$_3$ films and strain effects of buffer layers

Most epitaxial growths rely on the selection of proper substrates of the same structural symmetry as the acquired thin films. When the structural mismatch is substantially large, a trick commonly employed is to first deposit a buffer layer to partly alleviate the mismatch. An appropriate buffer layer is not only able to crystalize over the substrate itself, but have to own a crystal structure, in one way or another, lying between the structure of the substate and that of the thin film. (001)-oriented hexagonal manganite thin films can be epitaxially deposited both on sapphire (0001) and YSZ (111) substrate[2]. Here, we demonstrate that the buffer layer can be chosen as h-ReFeO$_3$ and h-ReMnO$_3$ to stabilize Lu$_{1-x}$Ca$_x$MnO$_3$ in hexagonal structure on rhombohedral sapphire substrates and cubic YSZ substrates, respectively. We have managed to grow h-Lu$_{1-x}$Ca$_x$MnO$_3$ thin films on YSZ substrate by intercalating a h-YMnO$_3$ buffer layer, as shown by XRD scans in Fig.S2(a). Because YSZ substrates turn to apply a tensile strain over all h-ReMnO$_4$ films [1,3], and the buffer layer h-YMnO$_3$ we use is only of about 1 nm, the Lu$_{1-x}$Ca$_x$MnO$_3$ thin films grown on top of them are expected to still be subject to the tensile strain from the YSZ substrate, only a small part of which might be relieved by h-h-YMnO$_3$. As one can see in Fig.S2(b), this is exactly verified by the thickness dependence of out-of-plane lattice constant c which increases as $t_{LC3MO}$ increases.

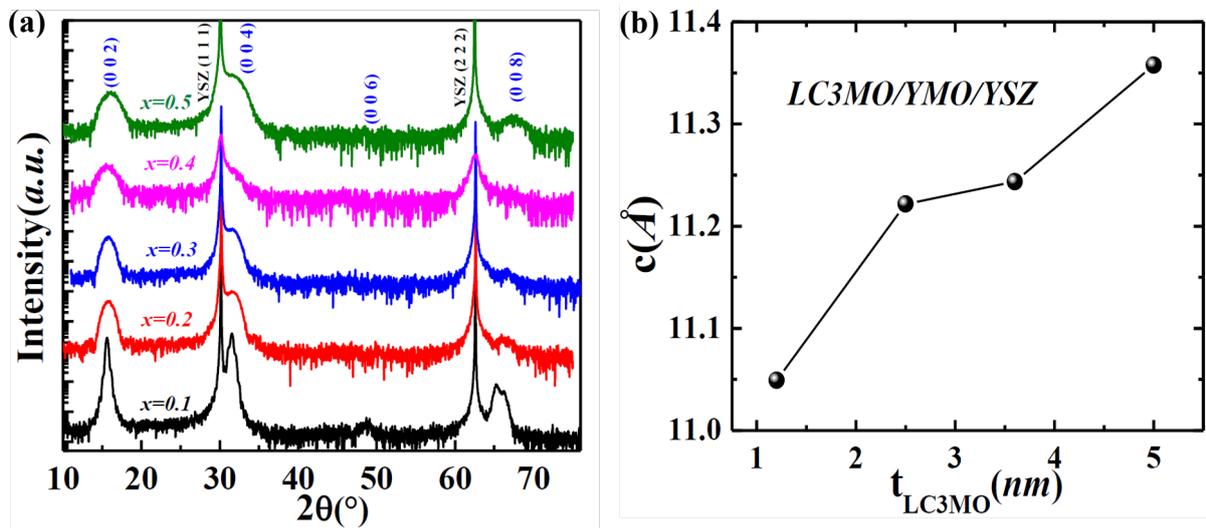

Fig.S2. **(a)** $\theta$-$2\theta$ scans of XRD for h-Lu$_{1-x}$Ca$_x$MnO$_3$ (001)/h-YMnO$_3$(001)/YSZ(111) samples with x=0.1(24 nm), 0.2 (5 nm), 0.3 (5 nm), 0.4(5 nm), 0.5 (5 nm). **(b)** Thickness $t_{LC3MO}$ dependent out-of-plane lattice constant c of h-L$_{0.7}$Ca$_{0.3}$MnO$_3$ films grown with the h-YMnO$_3$ buffer layer; the lattice constant value was derived from (0 0 2) peaks of the samples. Here, YMO and LC3MO denote h-YMnO$_3$, and h-Lu$_{0.7}$Ca$_{0.3}$MnO$_3$, respectively.

## 3. Temperature-resolved RHEED patterns of h-Lu$_{1-x}$Ca$_x$MnO$_3$/h-ScFeO$_3$/ $\alpha$-Al$_2$O$_3$ and LuMnO$_3$/YSZ Films along the [1 0 0] direction at critical thicknesses

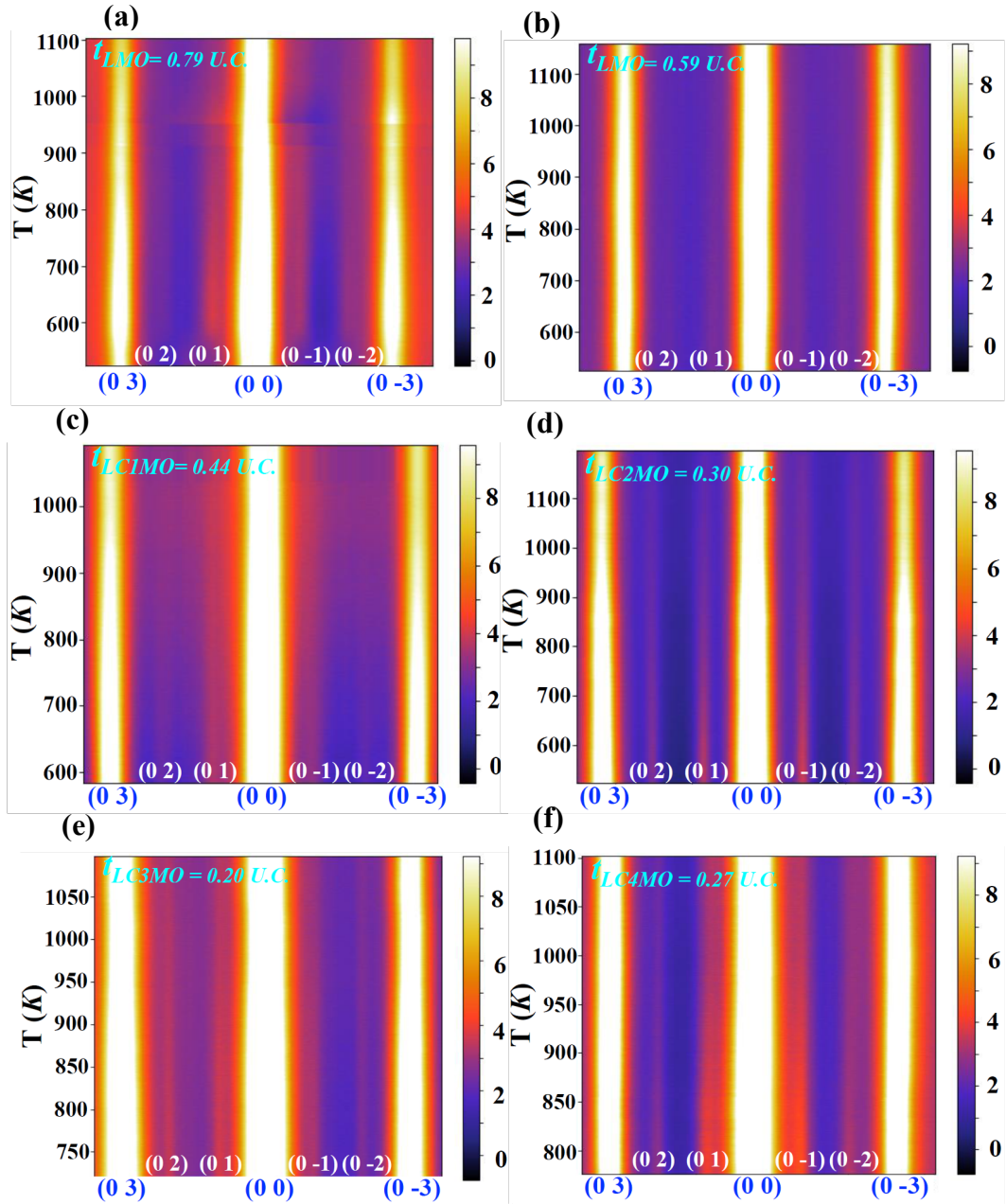

Fig.S3. Temperature-resolved RHEED patterns along the [1 0 0] direction at critical thicknesses for **(a)** h-LuMnO$_3$/YSZ; **(b)** h-LuMnO$_3$ /h-ScFeO$_3$/ $\alpha$-Al$_2$O$_3$ ; **(c)** h-Lu$_{0.9}$Ca$_{0.1}$MnO$_3$ /h-ScFeO$_3$/ $\alpha$-Al$_2$O$_3$ ; **(d)** h-Lu$_{0.8}$Ca$_{0.2}$MnO$_3$/h-ScFeO$_3$/$\alpha$-Al$_2$O$_3$; **(e)** h-Lu$_{0.7}$Ca$_{0.3}$MnO$_3$ /h-ScFeO$_3$/ $\alpha$-Al$_2$O$_3$; **(f)** h-Lu$_{0.6}$Ca$_{0.4}$MnO$_3$/h-ScFeO$_3$/ $\alpha$-Al$_2$O$_3$. Here, similarly, LMO, LC1MO, LC2MO, LC3MO and LC4MO stand for h-LuMnO$_3$, h-Lu$_{0.9}$Ca$_{0.1}$MnO$_3$, h-Lu$_{0.8}$Ca$_{0.2}$MnO$_3$, h-Lu$_{0.7}$Ca$_{0.3}$MnO$_3$, h-Lu$_{0.6}$Ca$_{0.4}$MnO$_3$, respectively. And $t_{LMO}$, $t_{LC1MO}$, $t_{LC2MO}$, $t_{LC3MO}$ and $t_{LC4MO}$ labeled on the top left corner of every diagram denote the samples' thicknesses. U.C. stands for unit cell.